\documentstyle[11pt,newpasp,twoside,epsf]{article}
\def\edcomment#1{\iffalse\marginpar{\raggedright\sl#1\/}\else\relax\fi}
\reversemarginpar

\begin{document}
\title{Information recovery from overlapping GAIA spectra}
 \author{Toma\v{z} Zwitter}
\affil{University of Ljubljana, Dept.\ of Physics, Jadranska 19,
1000 Ljubljana, Slovenia, tomaz.zwitter@fmf.uni-lj.si}

\begin{abstract}
The RVS instrument aboard GAIA is a slitless spectrograph, so some 
spectral tracing overlap is inevitable. We show that this background
can be accurately modeled and subtracted. Radial velocity accuracy is 
not degraded significantly unless the star density is higher than the 
value typical for the Galactic plane.
\end{abstract}

\section{Introduction}
GAIA RVS is a slitless spectrograph, so some degree of spectrum overlap is 
always present (Zwitter \&\ Henden 2002). The overlap is proportional to
resolution, but even low resolutions ($R \sim 5000$) do not avoid it. On
the other hand most astrophysical information other than radial velocity 
is washed away at resolutions lower than $R=10000$ 
(e.g. Bono 2002; Thevenin 2002). Overlapped stars are different during each transit,
so some spectra of the same object suffer from bad overlaps, others do not.
For this purpose it is essential that sidereal spin and precession periods of 
the satellite are not commeasurable.

Information recovery from overlapped spectra can use information from other
GAIA instruments. In particular, the astrometric position measurements
tell which stars are overlapping at each of the transits of a
given field over the focal plane. Also the distance between spectra heads  
(that could be treated as a rough ``velocity'' shift) is accurately known from the
star mappers in the spectro plane. 

Photometric classification of stellar spectra allows to choose correct spectral
templates for all overlapping spectra. Expected errors will be small, of the 
order of 100~K in temperature, 0.2 in $\log g$ and similar in metallicity. 
This information will build up during the mission, so reduction reiterations 
will be needed. 

\section{Radial velocity measurement}

Radial velocity supplies the sixth component in the position-velocity space, 
so it is the prime reason for existence of a spectrograph aboard GAIA. We 
ran a set of simulations to recover radial velocity. The results allowing 
only for zodiacal light background 
and read-out noise were published in Zwitter (2002), up-to-date results  
are discussed in Munari (2002).

%

Zodiacal light and read-out noise are not the only source of background. 
Spectra of other stars (partly) overlap our spectrum, depending on star 
density and spectral resolution (Zwitter \&\ Henden 2002). It is important 
that we know the positions, magnitudes and rough spectral types of 
overlapping background stars. We assume that:
\begin{itemize}
\item
Stellar positions are accurately known. Star mappers as well as astrometry
easily justify this assumption.
\item
Stellar magnitudes are accurately known. Typical errors will be 
$\sim 0.001$~mag for bright stars and $\le 0.02$~mag for stars of $V=18$
(ESA-SCI 2000(4)). This is well below the spectroscopic shot noise, so this 
assumption is justified. Contemporaneous star mapper flux measurements can 
supply the required information in case of variable stars. 
\item
Stellar types are roughly known. A mismatch of 250~K in temperature and 
0.5 in $\log g$ and [Fe$/$H] was assumed. This is some 3-times larger than 
expected typical final-mission stellar classification errors from 
photometry (Jordi 2002). So this assumption is justified even for 
mid-mission analysis. Simulated star types clustered around K1~V, 
i.e.\ a typical spectral type of background stars, and their luminosity 
function followed results of Zwitter \&\ Henden (2002). 
\item
Stellar radial velocities are roughly known. The assumed errors were
\begin{equation}
\sigma(v_r) = v_o \, \, \, 2.51^{V-14.0} 
\, \, \, \, \, \, , \, v_o = 0.2\,{\mathrm km/s}
\end{equation}
where $\sigma(v_r)$ is the standard deviation of the difference between 
assumed and true radial velocities and $V$ is the visual magnitude of the 
star. This is compatible to mission averaged results for a K1~V star
(Zwitter 2002). It turns out that radial velocity spread of background 
stars is not critical for the final results, so this assumption is 
justified. 
\end{itemize}

\begin{figure}
\plotfiddle{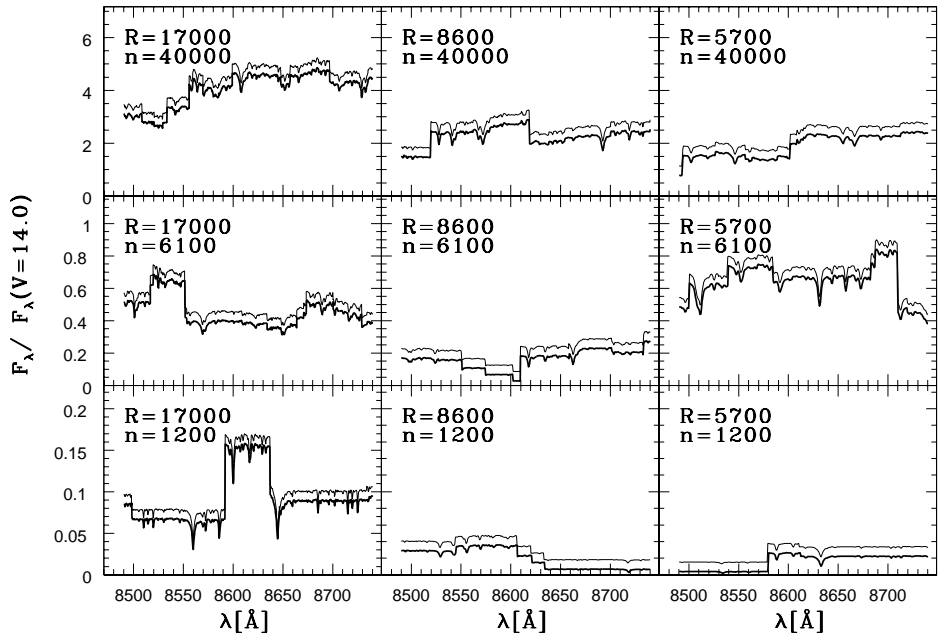}{7.9cm}{270}{48}{48}{-190}{270}
\caption{Examples of background spectra of different resolutions ($R$) 
and star densities ($n$). Thick lines: actual spectra. Thin lines: their
reconstruction from available information. A small vertical shift was 
applied to the latter for better legibility.}
\end{figure}

Figure 1 shows examples of background spectra. Vertical axis is spectral
flux in units of continuum flux from a K1~V star of magnitude $V=14$. Thus
the level of 0.01 corresponds to a contribution of a $V=19$ background
star. Randomly chosen examples of spectra at different resolutions ($R$)
and star densities ($n$) are plotted. Star density $n=1200$ ($V<17$) stars
per square degree is common at high Galactic latitudes ($|b| > 20^o$),
the value of 6100 is typical for the Galactic plane, while $n=40000$ 
($V<17$) stars/deg$^2$ is representative of rather high stellar densities,
encountered in $\le 10$\%\ of cases at $|b| < 20^o$ (Zwitter \&\ Henden 2002).
Number of overlapping background spectra and their flux level generally 
increases with resolution. 

One should not conclude that these rather bright and jumpy background 
spectra jeopardize derivation of radial velocities. 
The background can be very well modeled from information that 
is available (positions, magnitudes, rough spectral types and velocities of 
overlapping stars - see above). Note that this info yields spectral 
tracings (thin lines in Fig.\ 1) that are {\it very\/} similar to those of real 
stars (thick lines). 

Figure 2 reports final results for different star densities. Note that 
errors on derived radial velocity do not increase significantly, 
except for the faintest targets ($V>17$) and the highest star densities 
($n = 40000$ ($V<17$) stars per square degree). Even those errors could be 
reduced by filtering spectra so that only regions of spectral lines are 
cross-correlated. Curves are a bit noisy due to a limited Monte Carlo 
computing time, but the main result is clear: {\it Spectral lines of 
background stars do not spoil radial velocity analysis. The background 
stars merely increase the level of the background, thus increasing its 
shot noise. }

\begin{figure}
\plotfiddle{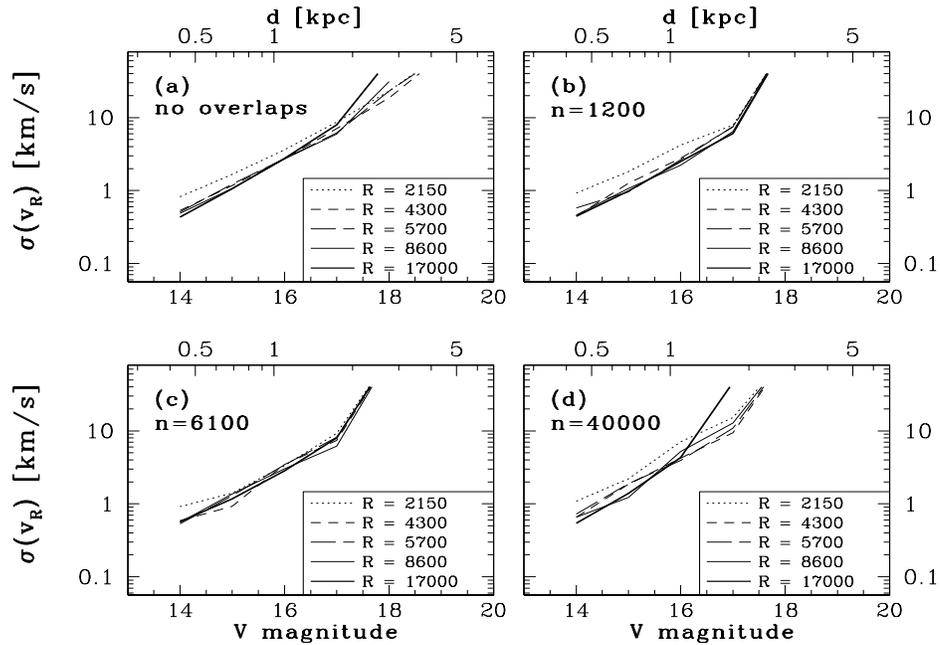}{7.9cm}{270}{49}{45}{-185}{257}
\caption{Radial velocity errors for a mission averaged spectrum of a 
K1~V star at different resolutions $R$. (a) no spectral overlaps, (b) observations at 
star densities characteristic for high Galactic latitudes,  
(c) for the Galactic plane, and (d) for a high density environment.
}
\end{figure}

\section{Recovery of other astrophysically relevant information}

GAIA spectrograph will be able to measure other information than radial 
velocities: abundances of individual elements, rotational velocity, 
temperature, $\log g$, and metallicity (see Thevenin 2002, Gomboc 2002). 
All these will be recovered from 
detailed spectral profiles of individual lines. For this purpose:
(a) spectral line profiles should be measured with high-enough accuracy;
(b) the profile should not be spoiled by notable spectral lines from 
overlapping background stars.

The first condition effectively limits the analysis to bright targets, 
$V < 15$ and the second to environments of moderate star density, $n<6000$ 
($V<17$) stars per square degree.

\section{Conclusions}

Radial velocities can be recovered even for faint targets. Because the spectral
shape of the background can be modeled accurately, overlapping stars degrade
the RV accuracy mainly by increasing the background shot noise. It was found
that overlapping spectra generally do not reduce the radial velocity accuracy 
substantially, even if observing at high resolution ($R \sim 10000$) and 
close to the Galactic plane. Degrading becomes substantial if the $S/N$ per
1~\AA\ bin of the non-overlapped spectrum is $<5$. This happens at $V \sim 17.5$
for mission averaged spectra and at $V \sim 15$ for single transits. Recovery 
of other astrophysical information is more difficult and mostly limited to 
bright targets in not too dense environments. For accurate modeling of the 
background the quality of stellar models as well as an accurate knowledge of flux 
throughput of the instrument (and filters used to bracket the spectral range)
is of utmost importance.

\end{document}